\documentclass[%
 aip,
 jmp,%
 amsmath,amssymb,figure*,
 reprint,%
]{revtex4-1}

\usepackage{graphicx}
\usepackage{dcolumn}
\usepackage{bm}

\begin{document}

\preprint{AIP/123-QED}

\title[Kaya et al.]{Nucleation and high-density packing of $360^{\circ}$ domain walls on planar ferromagnetic nanowires by using circular magnetic fields}

\author{F. I. Kaya}
\author{A. Sarella}%

\affiliation{ 
Department of Physics, Mount Holyoke College, South Hadley, MA, 01075, USA
}%

\author{D. Wang}
\author{M. Tuominen}
\affiliation{%
Department of Physics, University of Massachusetts, Amherst, MA, 01003, USA
}%
\author{K. E. Aidala}%
 \email{kaidala@mtholyoke.edu.}
\affiliation{ 
Department of Physics, Mount Holyoke College, South Hadley, MA, 01075, USA
}%

\date{\today}

\begin{abstract}
We propose a mechanism for nucleation and high-density packing of $360^{\circ}$ domain walls (DWs) on planar ferromagnetic nanowires, of $100$ nm width, by using circular magnetic fields. The extent of the stray field from a $360^{\circ}$ DW is limited in comparison to $180^{\circ}$ DWs,  which allows them to be packed more densely than $180^{\circ}$ DWs in a potential data storage device. We use micromagnetic simulations to demonstrate high-density packing of $360^{\circ}$ DWs, using a series of rectangular $16\times 16$ nm$^{2}$ notches to act as local pinning sites on the nanowires. For these notches, the minimum spacing between the DWs is $240$ nm, corresponding to a $360^{\circ}$ DW packing density of $4$ DWs per micron.  Understanding the topological properties of the $360^{\circ}$ DWs allows us to understand their formation and annihilation in the proposed geometry. Adjacent $360^{\circ}$ DWs have opposite circulation, and closer spacing result in the adjacent walls breaking into $180^{\circ}$ DWs and annihilating.  
%
\end{abstract}

\maketitle




Manipulating magnetic domain walls (DWs) in patterned ferromagnetic nanostructures and understanding their behavior are necessary to achieve proposed logic\cite{Hrkac} and data storage devices.\cite{sspparkin} Racetrack memory proposes the use of current driven transverse $180^{\circ}$ DWs, which interact over a range of about $2.5$ $\mu$m.\cite{Hayashi,Thomas} In contrast, $360^{\circ}$ DWs form an almost closed flux magnetic state, substantially reducing the interaction between neighboring DWs. For this reason, $360^{\circ}$ DWs have been proposed to serve as bits for data storage in a magnetic racetrack device.\cite{oyarce1}  A $360^{\circ}$ DW can be viewed as consisting of two $180^{\circ}$ DWs, and whether bringing together two transverse  $180^{\circ}$ DWs results in annihilation or a $360^{\circ}$ DW depends on the topological edge charges of the $180^{\circ}$ DWs.\cite{Pushp,Kunz1} Current driven motion of $360^{\circ}$ DWs is predicted to be different from $180^{\circ}$ DWs,\cite{Diegel,Mascaro} but experimental confirmation has been challenging.  Reliable nucleation and manipulation mechanisms are needed to study the properties of the $360^{\circ}$ DWs and to develop devices. Most proposals involve an injection pad with a rotating in-plane field,\cite{Diegel,Jang,Geng, Chen} with the exception of Gonzalez Oyarce et al.\cite{Oyarce} Here, we propose a versatile technique to controllably nucleate $360^{\circ}$ DWs  at arbitrary locations using a circular field centered in close proximity to a planar nanowire, allowing for the study of $360^{\circ}$ DWs in a wire and the potential to develop novel storage devices.

We perform micromagnetic simulations using the OOMMF\cite{OOMMF} package to iteratively solve the Landau-Lifshitz-Gilbert equation. The nanowire dimensions used in the simulations are $10000\times 100$ nm$^{2}$ and the material parameters are for permalloy: $M_{s}=8\times 10^{5}$ A/m, $A=1.3\times 10^{-11}$ J/m. The cell size is $4$ nm along the three axes, there is no crystalline anisotropy, the damping parameter is 0.5, and simulations are run at $0$ K. The magnetization state evolves until structures reach an equilibrium state where  $\frac{d\textbf{M}}{dt}<0.1$ deg/ns.

\begin{figure}
\centering
\includegraphics[scale=0.74]{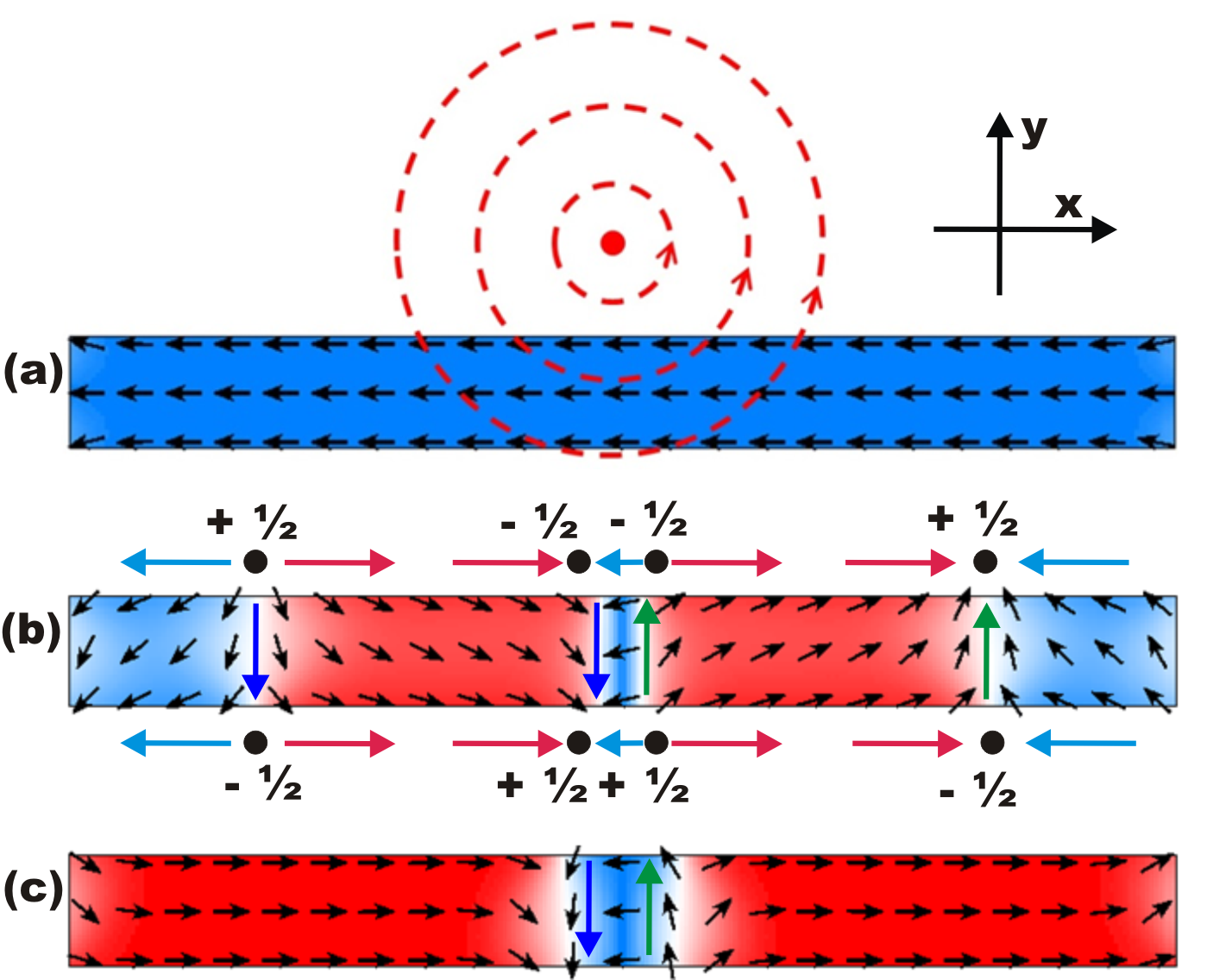}
\caption{
\label{fig:one}(a) Initialization of the nanowire and top-down view of the circular field. (b) Snapshot showing the nucleation of a $360^{\circ}$ DW and two $180^{\circ}$ DWs. (c) Relaxed state of a single $360^{\circ}$ DW. The color scale in (b) and (c) indicate the orientation of the moments along the $x$-axis: Red points to the right, blue to the left.  Green and blue arrows help identify the topological winding of the DWs.}
\end{figure}

Figure~\ref{fig:one}a shows the mechanism for nucleating a $360^{\circ}$ DW. We initialize the nanowire by magnetizing it along the negative $x$-axis with a large in-plane magnetic field of $160~mT$ or higher. We apply a circular field, simulated as if from a current in an infinitely long wire that flows into the page, which produces a field that decreases as $1/r$, where $r$ is the distance from the center of the field. A current of $21$ mA, which is located at a distance of $r = 48$~nm from the nanowire along the positive $y$-axis, corresponds to a field of $87.5$ mT at the top edge of the wire. The circular field exerts a torque on the magnetic moments, whereby it nucleates a $360^{\circ}$ DW in the nanowire directly below the center of the circular field (Fig.~\ref{fig:one}b). Two $180^{\circ}$ DWs are created on either side of the $360^{\circ}$ DW. The moments directly below the center of the magnetic field experience the smallest torque (theoretically zero in a perfect structure at zero temperature, since they are aligned opposite to the applied field), while the other moments feel stronger torques to align with the field. Such circular fields can be experimentally implemented via the tip of an Atomic Force Microscope (AFM),\cite{Goldman,Yang} 
which can be positioned at arbitrary locations to follow the pattern of fields described in this paper. For a scalable device, the procedure would presumably be realized by fabricating wires above each notch and by passing current through those wires perpendicular to the plane of the nanowire.

The simultaneous nucleation of two $180^{\circ}$ DWs on either side of the $360^{\circ}$ DW is a topological consequence, and is described in Bickel et al. for rings.\cite{Bickel} We characterize the $180^{\circ}$ DWs as ``up" or ``down" as conveniently revealed by whether the moments at their center are pointing in positive or negative y, indicated by the green or blue arrows in Figure $1b$. We use the same terminology for the $360^{\circ}$ DWs, which can be    ``up-down" or ``down-up" depending on the constituent $180^{\circ}$ DWs (read from left to right). Figure~\ref{fig:one}b is a snapshot in time during the nucleation of a $360^{\circ}$ DW, while the circular field is still applied. The half integer winding numbers of the topological edge charges\cite{Pushp,Kunz1} are indicated as well. At the nucleation of the $360^{\circ}$ DW, two topological defects with charge $-1/2$ are created on the top edge of the nanowire (Fig.~\ref{fig:one}b), and two $+1/2$ charges are created at the bottom. Two switched (red) domains appear on either side of the $360^{\circ}$ DW, aligning with the applied field. Two $180^{\circ}$ DWs must also be created (at the farther edge of the switched domain), and these must carry the opposite topological charges, $+1/2$ on the top and $-1/2$ on the bottom. The total winding number of the wire is zero, as required.

Given our CCW field and the resulting down-up $360^{\circ}$ DW, the $180^{\circ}$ DW that emerges to the right of the $360^{\circ}$ DW is an up $180^{\circ}$ DW, while the one to the left is a down $180^{\circ}$ DW. If a down $180^{\circ}$ DW joins with another down $180^{\circ}$ DW, the topological edge charges sum up to zero on the top and the bottom, hence the DWs annihilate. Similarly, the joining of two up $180^{\circ}$ DWs results in annihilation.

When the applied circular field is removed, the wire in Figure~\ref{fig:one} relaxes to the state shown in Figure~\ref{fig:one}c. The $180^{\circ}$ DWs are pushed to the side until they encounter the end of the wire and annihilate. This is generally not the case for longer wires in which the $360^{\circ}$ DW slides towards one of the $180^{\circ}$ DWs and eventually annihilates into a single $180^{\circ}$ DW as a result of the summation of the topological charges.

\begin{figure*}
\centering
\includegraphics[scale=0.66]{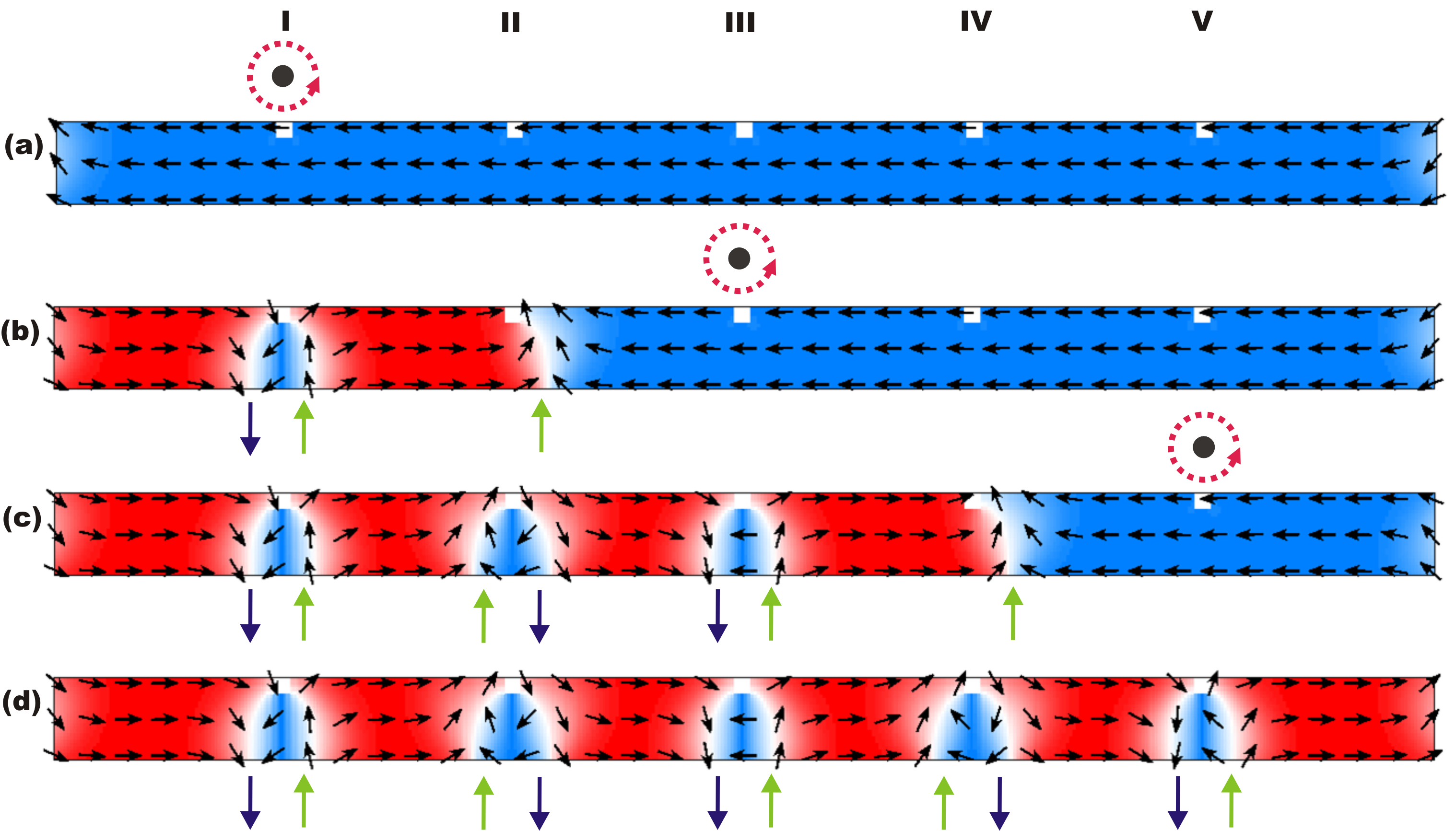}
\caption{
\label{fig:two} The sequence of steps for packing $360^{\circ}$ DWs of opposite circulation at adjacent notches. Red doted lines indicate the center of the CCW circular field.  (a) Initialization.  (b) Resulting state after applying $21$~mA above notch \textbf{I}.  (c) Resulting state after applying $21$~mA above notch \textbf{III}, creating a second $360^{\circ}$ DW directly below, and two $180^{\circ}$ DWs, one of which joins the $180^{\circ}$ DW at notch \textbf{II} to form a $360^{\circ}$ DW.  (d) Resulting state after applying  $21$~mA above notch \textbf{V}.  }
\end{figure*}

In order to pin $360^{\circ}$ DWs on the nanowire, a series of rectangular notches of $16\times 16$ nm$^{2}$ are introduced with an inter-notch distance of $280$ nm (Fig.~\ref{fig:two}). The length ($y$-axis) of the DW is reduced at the notches, thereby reducing the energy of the DWs and facilitating pinning at the notches. Figure $2$ demonstrates the sequence of steps required to generate a series of $360^{\circ}$ DWs with opposite circulation at adjacent notches. Once the nanowire is saturated along the negative $x$-axis as shown in Figure~\ref{fig:two}a, the first CW $360^{\circ}$ DW is nucleated at notch \textbf{I} by passing a current of $21$ mA vertically above the notch. As a result, an up $180^{\circ}$ DW pins at notch \textbf{II}, while the down $180^{\circ}$ DW slides to the end of the wire and annihilates due to the field gradient at the edge. The second $360^{\circ}$ DW at notch \textbf{III} is injected by following the same procedure.  The simultaneously nucleated down $180^{\circ}$ DW to the left pairs with the up $180^{\circ}$ DW that was earlier nucleated and pinned at notch \textbf{II}, forming a CCW $360^{\circ}$ DW as shown in Figure~\ref{fig:two}c. Similarly, the circular field is positioned at notch \textbf{V} to nucleate the CW $360^{\circ}$ DW at \textbf{V} and form the CCW $360^{\circ}$ DW at \textbf{IV}.  The magnetization circulation of the packed domain walls at notches \textbf{I} to \textbf{V} alternate between CW and CCW circulation, as shown in Figure~\ref{fig:two}d.

We have successfully simulated packing of $360^{\circ}$ DWs at adjacent notches with $260$ nm and $240$ nm inter-notch distances. As the notches are spaced more closely, the field strength is higher at notches adjacent to where the $360^{\circ}$ DW is nucleated. The $180^{\circ}$ DWs do not pin at the adjacent notch but are instead pinned two notches away. The second nucleated $360^{\circ}$ DW  must also be formed an additional notch away. It is straightforward to push these nucleated DWs to neighboring notches, by effectively unravelling the $360^{\circ}$ into two $180^{\circ}$ DWs with the correct strength field, and then pushing the $180^{\circ}$ DWs with an appropriate strength field.

\begin{figure}
\centering
\includegraphics[scale=0.66]{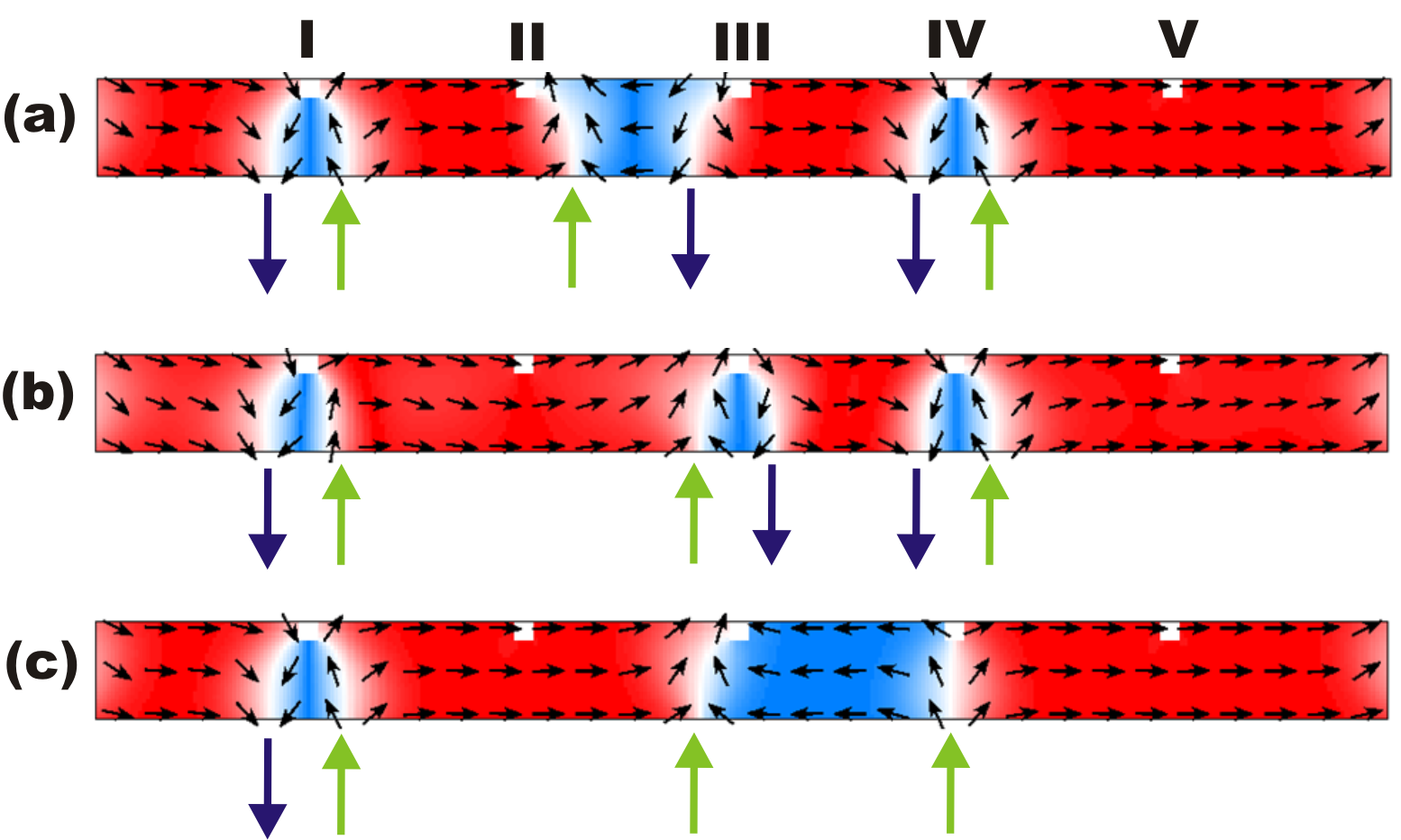}
\caption{
\label{fig:three} Failure mechanism when packing opposite circulation $360^{\circ}$ DWs at the inter-notch distance of $220$~nm.  (a) $360^{\circ}$ DWs are formed at notches \textbf{I} and \textbf{IV}.  The $180^{\circ}$ DWs interact but do not come together at a single notch.  (b) Temporarily applying a CCW field above notch \textbf{II} pushes the two $180^{\circ}$ DWs into a tight $360^{\circ}$ DW at notch \textbf{III}.  (c) When the field is removed, the constituent down DWs are sufficiently close to interact and annihilate, leaving the two up DWs at their respective notches. }
\end{figure}

The packing collapses if the distance between adjacent notches is $\leq 220$~nm. Figure~\ref{fig:three} shows why the previously described procedure fails when the notches are too close together, at $220$ nm. We first nucleate the down-up $360^{\circ}$ DW at \textbf{I}, creating an up $180^{\circ}$ DW at \textbf{II}. We nucleate the second $360^{\circ}$ DW at notch \textbf{IV}, as the $180^{\circ}$ DW at notch \textbf{II} is too close to notch \textbf{III} and prevents the nucleation of a down $180^{\circ}$ DW to the left of notch \textbf{III}. We instead nucleate the down-up $360^{\circ}$ DW at notch \textbf{IV}, and see that the down $180^{\circ}$ DW moves to notch \textbf{III} and the $180^{\circ}$ DWs at \textbf{II} and \textbf{III} are interacting, shown in Figure~\ref{fig:three}a.  Figure~\ref{fig:three}b shows that by applying a CCW field at notch \textbf{III}, we can temporarily form a tight $360^{\circ}$ DW pinned at notch \textbf{III}.  However, once the field is removed (Fig.~\ref{fig:three}c),  the $360^{\circ}$ DWs at notch \textbf{III} and \textbf{IV} annihilate one another due to their topological charges. Effectively, the two down constituent $180^{\circ}$ DWs are adjacent, attract each other, and annihilate. The two up $180^{\circ}$ DWs remain at notches \textbf{III} and \textbf{IV}. Therefore, a distance of $\leq 220$ nm between notches prevents packing of $360^{\circ}$ DWs at adjacent notches on a nanowire, using this technique and geometry.

The minimum spacing between $360^{\circ}$ DWs is determined in part by the notch size and shape.  Deeper notches allow closer packing but require stronger fields to de-pin the DWs. We have succeeded in simulating a dense packing of $360^{\circ}$ DWs at adjacent notches with $220$ nm spacing by using $16\times 32$ nm$^{2}$ rectangular notches. The procedure in this case differs slightly due to the stronger pinning of $180^{\circ}$ and $360^{\circ}$ DWs at deeper notches. Additionally, if we control the topology of the adjacent $360^{\circ}$ DWs so that they are all of the same circulation, the failure mechanism changes and we can pack the $360^{\circ}$ DWs more densely. This can be accomplished by  annihilating the DW with the circulation that we do not want by using a strong local field above that DW. For example, a strong enough ($85$~mA) CCW field at notch I in Figure~\ref{fig:three}a annihilates the  $360^{\circ}$ DW pinned at \textbf{I}.  We can then shift the other $360^{\circ}$ DWs by unravelling them into two $180^{\circ}$ DWs and pushing the $180^{\circ}$ DWs. For $16\times 64$ nm$^{2}$ rectangular notches, we can successfully pack $360^{\circ}$ DWs with the opposite circulation at $180$~nm spacing between the notches. More work remains to be done to better understand the effects of the geometry of the notches and the circulation of adjacent $360^{\circ}$ DWs  and their effects on the packing density.\cite{Prep}

While these notched wires allow us to study the behavior of $360^{\circ}$ DWs, and using the tip of an AFM to manipulate the DWs provides flexibility in our experiments, a realistic device would have prefabricated wires positioned above each notch where we center the circular field in our simulations. The presence or absence of the $360^{\circ}$ DW could be used as the bit, or possibly the circulation of the $360^{\circ}$ DW. Geometry would be optimized to reduce the current density and power consumption while maintaining a close packing density. The readout might be similar to racetrack memory,\cite{Hayashi,Thomas} requiring a spin-polarized current to move the $360^{\circ}$ DWs. Generally, there will be a trade-off between strong pinning providing closer packing, and weak pinning requiring smaller fields and current to move the DWs.

In summary, we propose a mechanism to nucleate $360^{\circ}$ DWs at arbitrary locations determined by notches along an in-plane ferromagnetic nanowire. A circular field that decreases as $1/r$ and is centered directly above a notch along the y-axis will nucleate one $360^{\circ}$ DW and two $180^{\circ}$ DWs at that notch. Careful consideration of the series of circular fields allows us to nucleate $360^{\circ}$ DWs with opposite circulation at adjacent notches as close as $240$~nm, providing a packing density of about four DWs per micron in the permalloy nanowire simulated with $16\times 16$~nm$^{2}$ rectangular notches.\\

The authors acknowledged the support by NSF grants No. DMR 1208042 and 1207924. Simulations were performed with the computing facilities provided by the Center for Nanoscale Systems (CNS) at Harvard University (NSF award ECS-0335765), a member of the National Nanotechnology Infrastructure Network (NNIN).

\bibliography{aipsamp.bib}

\end{document}